\documentclass[aps,prb,showpacs,twocolumn,floats,epsfig,pdflatex]{revtex4}
\usepackage{amssymb}
\usepackage{amsbsy}
\usepackage{amsmath}
\usepackage{epsfig}
\usepackage{graphicx}
\begin{document}

\title {Ultracold bosons in a synthetic periodic magnetic field: Mott phases and
re-entrant superfluid-insulator transitions}

\author{K. Saha, K. Sengupta, and K. Ray}

\affiliation {Theoretical Physics Department, Indian Association for
the Cultivation of Science, Kolkata-700032, India. }

\date{\today}

\begin{abstract}

We study Mott phases and superfluid-insulator (SI) transitions of
ultracold bosonic atoms in a two-dimensional square optical lattice
at commensurate filling and in the presence of a synthetic periodic
vector potential characterized by a strength $p$ and a period
$l=qa$, where $q$ is an integer and $a$ is the lattice spacing. We
show that the Schr\"odinger equation for the non-interacting bosons
in the presence of such a periodic vector potential can be reduced
to an one-dimensional Harper-like equation which yields $q$ energy
bands. The lowest of these bands have either single or double minima
whose position within the magnetic Brillouin zone can be tuned by
varying $p$ for a given $q$. Using these energies and a
strong-coupling expansion technique, we compute the phase diagram of
these bosons in the presence of a deep optical lattice. We chart out
the $p$ and $q$ dependence of the momentum distribution of the
bosons in the Mott phases near the SI transitions and demonstrate
that the bosons exhibit several re-entrant field-induced SI
transitions for any fixed period $q$. We also predict that the
superfluid density of the resultant superfluid state near such a SI
transition has a periodicity $q$ ($q/2$) in real space for odd
(even) $q$ and suggest experiments to test our theory.

\end{abstract}


\maketitle

\section{Introduction}
\label{intro}

Several experiments on ultracold trapped atomic gases have opened a
new window onto the phases of quantum matter \cite{Greiner1}. A gas
of bosonic atoms in an optical or magnetic trap has been reversibly
tuned  between superfluid and insulating ground states by varying
the strength of a periodic potential produced by standing optical
waves \cite{Greiner1,Orzel1}. This transition has been explained on
the basis of the Bose-Hubbard model with on-site repulsive
interactions and hopping between nearest neighboring sites of the
lattice. \cite{fisher1,sachdev1, jaksch1,sesh1,dupuis1}. In fact,
experiments on the superfluid-insulator (SI) transitions of such
bosonic atoms in two-dimensional (2D) optical lattices
\cite{spielman1} is found to agree with predictions of theoretical
studies of the Bose-Hubbard model quite accurately
\cite{fisher1,dupuis1,freericks1}.

More recently, several experiments have successfully generated time-
or space- dependent effective vector potentials for neutral bosons.
Such synthetic vector potentials are created by generating
temporally or spatially dependent optical coupling between the
internal states of these bosonic atoms \cite{gaugepapers1,
spielman2, spielman3}. We note that this experimental technique
involves production of a specific effective vector potential for the
atoms and hence corresponds to a fixed gauge. In the simplest
experimental setup, these vector potentials are typically chosen to
represent a constant magnetic field in the asymmetric gauge.
However, a few experiments have also generated vector potentials
which correspond to spatially varying synthetic magnetic fields
\cite{spielman3}. Several theoretical studies have been carried on
the properties of the bosons in deep optical lattice in the presence
of a constant synthetic magnetic field \cite{ucpapers1}. In
particular, the SI phase boundary has been computed both using
mean-field theory \cite{lundhoker1} and excitation energy
calculation which relies on a perturbative expansion in the hopping
parameter \cite{freericks2}. More recently, experimentally relevant
issues, such as the momentum distribution of the bosons in the Mott
phase, the critical theory of the SI transition, and the nature of
the superfluid ground states and collective modes near criticality
have also been addressed \cite{sengupta1,sensarma1}. However, in
spite of the possibility of direct experimental realization
\cite{spielman3}, the phase diagram of these bosons in the presence
of a spatially dependent magnetic field has not been theoretically
investigated.

In this work, we present a theory of the SI transition for ultracold
bosons in a 2D square optical lattice with commensurate filling
$n_0$ and in the presence of a periodic synthetic vector potential
given by $\vec{A}^{\ast}=(0, A^{\ast}_y)$ with $A^{\ast}_y=
A_0^{\ast} \sin(2\pi x/l)$, where $l=qa$ is the period of the vector
potential, $q$ is an integer, $a$ is the lattice spacing, and
$A_0^{\ast}$ is the maximum value of the vector potential on any
lattice site. At the outset, we introduce a dimensionless number $p=
2 \pi q^{\ast} A_0^{\ast} a/hc$, (where $q^{\ast}$ is the effective
charge of the bosons \cite{spielman2}, $c$ is the speed of light,
and $h = 2 \pi \hbar$ is the Planck's constant) which will be used
in the rest of this work to characterize the strength of the vector
potential. We first consider the problem of non-interacting bosons
in a lattice in the presence of such a periodic vector potential and
show that the corresponding single particle Schr\"odinger equation
can be reduced to a one-dimensional Harper-like equation
\cite{hof1,koh1}. The solution of this equation yields an energy
spectra with $q$ bands (with energies $\epsilon_{\alpha}^q ({\bf
k};p)$ for $\alpha=0..q-1$) all of which have a periodicity of $2
\pi/q$ along $k_x$. The lowest of these bands $\epsilon_0^q({\bf
k};p)$ has, depending on $p$, either a single minimum at ${\bf
k}\equiv (k_x,k_y)=(0,0)$ or $(0,\pi)$ or doubly degenerate minima
either at $(0,0)$ and $(0,\pi)$ or at $(0, \pm k_y^{\rm min})$ where
$k_y^{\rm min}$ can vary continuously as a function of $p$ for a
given $q$. The minimum energy of the lowest band, $\epsilon_{\rm
min}$, turns out to be a non-monotonic function of $p$ for a fixed
$q$. Using these properties of the single particle energy bands and
a strong coupling expansion \cite{dupuis1,sengupta1}, we analyze the
Mott phase and SI phase transition of these bosons in the presence
of a deep optical lattice. We show that, depending on $p$ and $q$,
the momentum distribution of these bosons in the Mott phase near the
SI transition will exhibit single (double) precursor peak(s) at the
position of the minimum (minima) of $\epsilon_0^q({\bf k};p)$. We
determine the SI phase boundary and demonstrate that the bosons
exhibit a series of re-entrant field-induced SI transitions as a
function of the vector potential strength $p$ for any period $q$. We
also construct an effective Landau-Ginzburg action for the SI
transition and show, by analyzing this action at a mean-field level,
that the resultant superfluid state has a $q$ ($q/2$) periodic
structure in real space for any odd (even) $q$. We show that the
reason for such a period-halving of the superfluid density for even
$q$ can be traced back to the properties of the Harper-like equation
obeyed by the non-interacting bosons. We discuss several experiments
that can probe our theory.

The rest of the paper is organized as follows. In Sec.\ \ref{sp}, we
introduce the relevant tight-binding Hamiltonian of the bosons in an
optical lattice in the presence of the periodic vector potential and
obtain the energy spectrum when the interaction between these bosons
is set to zero. This is followed by Sec.\ \ref{sib}, where we
introduce the strong coupling expansion for the bosons and use it to
compute the boson momentum distribution in the Mott phase and the SI
phase boundary. In Sec.\ \ref{sf}, we show that the superfluid state
into which the transition takes place exhibits a $q$-periodic
superfluid density. We conclude with a discussion of possible
experiments to test our theory in Sec.\ \ref{conc}.

\section{Non-interacting Boson spectrum}
\label{sp}

The Hamiltonian of a system of bosons in the presence of an optical
lattice and a synthetic periodic vector field is given by
\cite{Greiner1, fisher1,spielman1,lundhoker1,freericks2}
\begin{eqnarray}
{\mathcal H} &=& \sum_{{\bf r},{\bf r'}} t'_{{\bf r}{\bf r'}}
b_{{\bf r}}^{\dagger} b_{{\bf r'}} + \sum_{{\bf r}} [-\mu {\hat
n}_{{\bf r}} + \frac{U}{2} {\hat n}_{{\bf r}}({\hat n}_{{\bf r}}-1)
] \label{ham1}
\end{eqnarray}
where $\mu$ is the chemical potential, $U$ is the on-site Hubbard
interaction, $b_{\bf r}$ (${\hat n}_{\bf r}= b_{\bf r}^{\dagger}
b_{\bf r}$) is the boson annihilation (density) operator, the
hopping matrix $t'_{{\bf r r'}}$ is given by
\begin{eqnarray}
t'_{{\bf r}{\bf r'}} &=&  -t' e^{-i q^{\ast} \int^{\bf r'}_{\bf r}
\vec{A^{\ast}} \cdot \vec{dl}/ \hbar c}, \label{hop1}
\end{eqnarray}
if ${\bf r}\equiv (x,y)=(m,n)a$ and ${\bf r'}$ are nearest
neighboring sites and is zero otherwise, and $t'$ is the hopping
amplitude of the bosons between the nearest neighboring sites. In
the rest of this work, we set the lattice spacing $a$, $\hbar$, and
$c$ to unity. Our aim in this work is to analyze the phases of
${\mathcal H}$.

To this end, we first analyze the boson spectrum in the
non-interacting limit $\mu=U=0$. In this case, non-interacting boson
Hamiltonian becomes
\begin{eqnarray}
{\mathcal H}_0 &=& -t' \sum_{m,n} \Big[ b_{mn}^{\dagger}
\left(b_{m+1,n} +
b_{m,n+1} e^{-ip \sin(2\pi m/q)} \right) \nonumber\\
&& + {\rm h.c} \Big], \label{hamreal}
\end{eqnarray}
where we have used $p= 2 \pi q^{\ast} A_0^{\ast} a/hc$. To obtain
the spectrum for $H_0$, we use the identity
\begin{eqnarray}
e^{iz\sin x} = \sum_{r=-\infty}^{\infty} J_{r}(z) e^{i x r},
\end{eqnarray}
where $J_r(z)$ denotes Bessel functions with integer $r$, and write
${\mathcal H_0}$ in momentum-space representation as
\begin{eqnarray}
{\mathcal H}_0 &=& -t' \sum_{{\bf k}} \Big[ 2 \cos(k_x) b^{\dagger}
(k_x,k_y) b(k_x,k_y)  \nonumber\\
&& + \Big( \sum_{r=-\infty}^{\infty}  J_r (p) e^{-i k_y} b^{\dagger}
(k_x,k_y) b( k_x + 2 \pi r/q, k_y) \nonumber\\
&& + {\rm h.c} \Big) \Big] \nonumber\\
&=& -t' \sum_{{\bf k}} \Big[ 2(\cos(k_x) +  S_0(p)
\cos(k_y)) \nonumber\\
&& \times b^{\dagger} (k_x,k_y) b(k_x,k_y) + \sum_{r=1}^{q-1}
S_{r}(p) e^{-i k_y} \nonumber\\
&& \times b^{\dagger} (k_x,k_y) b( k_x + 2 \pi r/q, k_y) + {\rm h.c}
\Big], \label{hamni}
\end{eqnarray}
where $ b(k_x,k_y) = \sum_{{\bf k}} \exp(i(k_x m + k_y n)) b_{mn}$.
In Eq.\ \ref{hamni}, $S_{r}(p)$ is given by
\begin{eqnarray}
S_{r}(p) = \sum_{n=-\infty}^{\infty} J_{qn+r}(p),
\end{eqnarray}
where $n$ takes integer values, $S_{r}(p) = S_{r+q}(p)$, and we have
used the $2 \pi$ periodicity of $b(k_x,k_y)$: $b(k_x+2
\pi,k_y)=b(k_x,k_y)$. Note that for even $q$, $S_{r}=0$ for all odd
integer $r$ which follows from the well-known property of the Bessel
functions $J_n(p)= (-1)^n J_{-n}(p)$ for any integer $n$.

\begin{figure}
\rotatebox{0}{\includegraphics*[width=\linewidth]{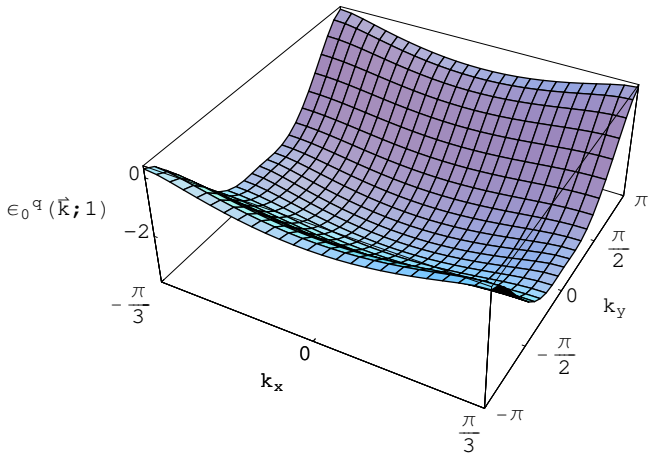}}
\caption{(Color online) Plot of $\epsilon_0^{(3)}(\vec k;1)$ showing
a single minima at $(k_x,k_y)=(0,0)$.} \label{fig1}
\end{figure}

The Schr\"odinger equation obtained from Eq.\ \ref{ham1} can be
written by expressing the eigenfunctions as \cite{koh1}
\begin{eqnarray}
|\psi\rangle = \sum_{\alpha=0}^{q-1} \psi_{\alpha} b^{\dagger}( k_x
+ 2 \pi \alpha/q, k_y) |0\rangle, \label{eigenvec}
\end{eqnarray}
where $\psi_{\alpha} = \psi_{\alpha+q}$, and obtaining the equations
of $\psi_{\alpha}$ from ${\mathcal H_0} |\psi\rangle = E
|\psi\rangle$. This yields a one-dimensional Harper-like equation
for $\psi_{\alpha}$
\begin{eqnarray}
\epsilon \psi_{\alpha} &=& -t' \Big[ 2(\cos(k_x) + S_0(p) \cos(k_y))
\psi_{\alpha} \nonumber\\
&& + \sum_{r=1}^{q-1} S_{r}(p) \left(e^{-i k_y} \psi_{\alpha+r} +
e^{i k_y} \psi_{\alpha-r} \right) \Big]. \label{sch1}
\end{eqnarray}
Eq.\ \ref{sch1} can easily be cast in the form of $q \times q$
dimensional Hermitian matrix equation $\Lambda^q({\bf k};p)
\psi=\epsilon \psi$. The diagonal elements of $\Lambda^q({\bf k};p)$
are given by $ \Lambda_{nn}^q({\bf k};p)= -2(\cos(k_x+2 \pi (n-1)/q)
+ S_0(p) \cos(k_y))$ and the off-diagonal elements by
$\Lambda_{n,n+r}^q({\bf k};p)= \Lambda_{n+r,n}^{q \ast} ({\bf k};p)
= -S_{r}(p) e^{-i k_y}$. The difference of $\Lambda^q({\bf k};p)$
with its counterpart in the constant magnetic field \cite{koh1} is
two-fold. First, $\Lambda^q({\bf k};p)$ no longer remains a
tri-diagonal matrix. However, the $2\pi/q$ periodicity of its
eigenvalues, which is a consequence of the periodicity of the
magnetic field, is still retained. This property is most easily seen
by noting that a shift of $k_x \to k_x + 2\pi/q$ in Eq.\
\ref{eigenvec} amounts to a shift of $\psi_{\alpha} \to
\psi_{\alpha+1}$. Second, for even $q$, where
$\Lambda_{n,n+r}^q({\bf k};p)= \Lambda_{n+r,n}^{q \ast} ({\bf k};p)
=0$ for all odd $r$, $\Lambda^q({\bf k};p)$ separates into two
block-diagonal matrices of dimension $q/2$ leading to $q/2$ non-zero
elements of the eigenvector $\psi$ for any eigenvalue $\epsilon$.
Note that for $q=2$, which correspond to $A_y^{\ast}=0$ on all
sites, we have $S_0(p)=1$ and $S_1(p)=0$ so that Eq.\ \ref{sch1}
reduces to the standard tight-binding Hamiltonian in zero magnetic
field.
\begin{figure}
\rotatebox{0}{\includegraphics*[width=\linewidth]{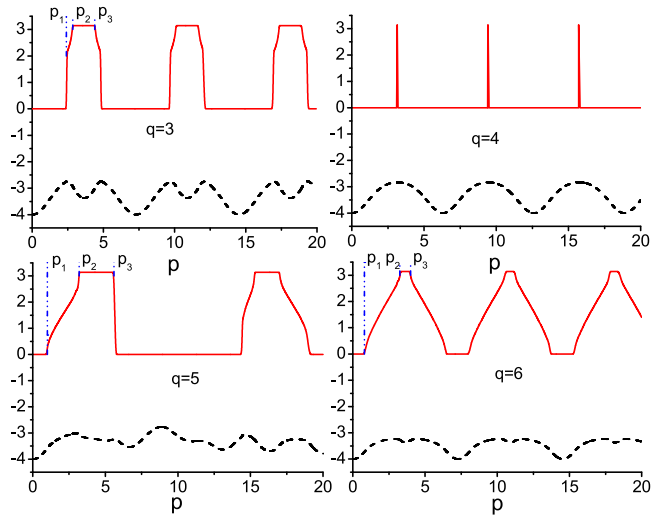}}
\caption{(Color online) Plot of $k_y^{\rm min}$ (red solid line) and
$\epsilon_{\rm min}/t'$ (black dashed line) as a function of the
vector potential strength $p$ for $q=3$ (left top panel), $4$ (right
top panel), $5$ (left bottom panel), and $6$ (right bottom panel),
showing non-monotonic periodic behavior. See text for details. }
\label{fig2}
\end{figure}

For $q \ge 3$, a straightforward numerical diagonalization of
$\Lambda$ leads to $q$ energy bands with energy dispersions
$\epsilon^q_{\alpha}({\bf k};p)$, where $\alpha=0..q-1$, which have
a period of $2 \pi/q$ along $k_x$. This periodicity is a
manifestation of the q-fold folding of the Brillouin zone due to the
presence of the periodic vector potential. The lowest energy band
$\epsilon_0^q({\bf k};p)$, shown in Fig.\ \ref{fig1} for $p=1$ and
$q=3$, displays a single minima at $(k_x,k_y)=(0,0)$ within the
magnetic Brillouin zone ($-\pi/q \le k_x \le \pi/q$ and $-\pi\le k_y
\le \pi$). This minima structure changes with increasing $p$ as
shown in Fig.\ \ref{fig2} for $q=3$, $4$, $5$ and $6$. For $q=3$,
$5$ and $6$, we find that beyond a critical strength of the vector
potential $p_1(q)$, $\epsilon_0^q({\bf k};p)$ has two minima at the
$(0, \pm k_y^{\rm min}(p))$. As $p$ is increased, $k_y^{\rm min}$
increases monotonically from $0$ to $\pi$ until it reaches $\pi$ at
$p=p_2(q)$, where we recover the single minima structure of
$\epsilon_0^q({\bf k};p)$ with the minima at $(0,\pi)$. As $p$ is
further increased, till a value $p_3(q)$, $k_y^{\rm min}$ remains at
$\pi$. Beyond $p_3(q)$, for $q=3$ and $6$, we find that
$\epsilon_0^q({\bf k},p)$ again has two minima at $(0,\pm k_y^{\rm
min}(p))$ and $k_y^{\rm min}(p)$ monotonically decreases from $\pi$
to $0$ as $p$ is increased. For $q=5$, beyond $p_3(q)$, we find a
discontinuous change in $k_y^{\rm min}$ from $\pi$ to $0$, and
$\epsilon_0^q({\bf k};p)$ retains its single minima structure. For
$q=4$, we always have a single minima of $\epsilon_0^q({\bf k};p)$
at $(k_x,k_y)=(0,0)$, except at $p=n \pi$ where there are two
degenerate minima at $(0,0)$ and $(0,\pi)$. We also note from Fig.\
\ref{fig2}, that the minimum value of the energy, $\epsilon_{\rm
min}$, is a non-monotonic function of $p$ for all $q\le 6$. We have
checked that these features remain qualitatively similar for $q>6$
and we shall not discuss those cases further here. In the next
section, we shall utilize these properties of $\epsilon_{0}^q({\bf
k};p)$ to understand the phase diagram of these bosons in the
presence of a deep optical lattice.

Before ending this section, we note that there is an alternative
method of finding the energy eigenvalues of the Hamiltonian Eq.\
\ref{hamreal} by constructing the Schr\"odinger equation in real
space and using the $q$ periodicity of the eigenfunctions along $x$.
This has been carried out in Ref.\ \onlinecite{oh1} and yields
identical results to the method elaborated here. We also point out
that, although we have, keeping in mind the simplicity of
experimental realization, considered a relatively simple sinusoidal
form of the vector potential, our method can be easily generalized
to treat more complicated periodic vector potentials. Also, we note
that since the vector potential $A_y^{\ast}$ is not a \emph{gauge
field}, there is no gauge freedom in the choice of the
eigenfunctions (Eq.\ \ref{eigenvec}). Thus the flux of the vector
potential appears only in the coefficients of the Harper equation
and not in the choice of the eigenfunctions which is in contrast to
the case of periodic magnetic fields with gauge freedom treated in
Ref.\ \onlinecite{gn1}.

\section{Strong Coupling Expansion}
\label{sib}

In this section, we analyze the phases of ${\mathcal H}$ in the
limit of $t'/U \ll 1$, where the bosons are in a Mott insulating
state. We note that the effect of the magnetic field manifests
itself in the first term of Eq.\ \ref{ham1} and thus vanishes in the
local limit ($t'=0$). In this limit the boson Green function can be
exactly computed \cite{dupuis1,sengupta1,freericks1} and is given,
at $T=0$, by
\begin{eqnarray}
G_0(i\omega_n) = \frac{(n_0+1)}{i \omega_n - E_p} - \frac{n_0}{i
\omega_n + E_h}. \label{greenlocal}
\end{eqnarray}
Here $\omega_n$ denote bosonic Matsubara frequencies and $E_h = \mu
-U(n_0-1)$($E_p=-\mu + Un_0$) are the energy cost of adding a hole
(particle) to the Mott state.
\begin{figure}
\rotatebox{0}{\includegraphics*[width=\linewidth]{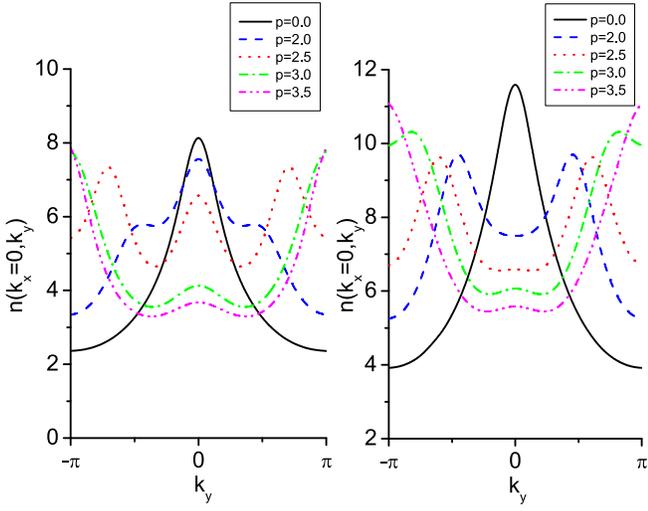}}
\caption{(Color online)Plot of $n(k_x=0,k_y)$ as a function of $k_y$
for different representative value of $p$, and for $q=3$ (panel $1$)
and $q=5$ (panel $2$) at $\mu/U=0.414$ and
$t'(p)/t'_c(p)=0.95$.}\label{fig3}
\end{figure}
To address the effects of the hopping term, we resort to the
coherent state path integral description of these bosons. The
partition function of the system can then be written as
\begin{eqnarray}
Z &=& \int D \tilde \psi D \tilde \psi^{\ast} e^{-(S_0 +S_1)}, \nonumber\\
S_0 &=& \int_0^{\beta} d\tau \sum_{{\bf r}} \Big[ \tilde
\psi^{\ast}_{{\bf r}}(\tau)
\partial_{\tau} \tilde \psi_{\bf r}(\tau) - \mu n_{\bf r}(\tau)
\nonumber\\
&& + \frac{U}{2} n_{\bf r}(\tau) (n_{\bf r}(\tau)-1)\Big], \nonumber\\
S_1 &=& \int_0^{\beta} d\tau \sum_{{\bf r},{\bf r'}} t'_{{\bf r}{\bf
r'}} {\tilde \psi}_{{\bf r}}^{\ast}(\tau)  {\tilde \psi}_{{\bf
r'}}(\tau). \label{ac1}
\end{eqnarray}
Here $\tau$ is the imaginary time, ${\tilde \psi}$ denote boson
fields in the path integral representation, $n_{{\bf r}}(\tau)=
{\tilde \psi}_{{\bf r}}^{\ast}(\tau)  {\tilde \psi}_{{\bf
r}}(\tau)$, $\beta=1/k_B T$ is the inverse temperature ($T$), and
$k_B$ is the Boltzman constant. Following Ref.\
\onlinecite{dupuis1}, we then decouple the hopping term introducing
a Hubbard-Stratonovitch field $\phi_{\bf r}(\tau)$. The partition
function can then be written as
\begin{eqnarray}
Z &=& \int D \tilde \psi D \tilde \psi^{\ast} D \phi D \phi^{\ast}
e^{-(S_0 +S'_1 +S'_2)}, \nonumber\\
S'_1 &=& \int_0^{\beta} d\tau \sum_{{\bf r}} (\tilde
\psi^{\ast}_{{\bf r}}(\tau) \phi_{\bf r}(\tau) + {\rm h.c}),
\nonumber\\
S'_2 &=& - \int_0^{\beta} d\tau \sum_{{\bf r},{\bf r'}}
t^{'-1}_{{\bf r}{\bf r'}} \phi_{{\bf r}}^{\ast}(\tau) \phi_{{\bf
r'}}(\tau). \label{ac2}
\end{eqnarray}
Finally, we introduce a second Hubbard-Stratonovitch field
$\psi_{\bf r}(\tau)$ and decouple $S'_2$ to obtain
\begin{eqnarray}
Z &=& \int D \tilde \psi D \tilde \psi^{\ast} D \phi D \phi^{\ast}
D \psi D \psi^{\ast}  e^{-(S_0 +S'_3 +S'_4)}, \nonumber\\
S'_3 &=& \int_0^{\beta} d\tau \sum_{{\bf r}} \left[ (\tilde
\psi^{\ast}_{{\bf r}}(\tau)- \psi_{\bf r}^{\ast} (\tau)) \phi_{\bf
r}(\tau) + {\rm h.c} \right],
\nonumber\\
S'_4 &=& \int_0^{\beta} d\tau \sum_{{\bf r},{\bf r'}} t'_{{\bf
r}{\bf r'}} \psi_{{\bf r}}^{\ast}(\tau) \psi_{{\bf r'}}(\tau).
\label{ac3}
\end{eqnarray}
Note that integrating out $\phi_r(\tau)$ in Eq.\ \ref{ac3} would
lead to the constraint $\psi_r = {\tilde \psi}_r$ on $Z$. It can
also be shown that $\psi$ and ${\tilde \psi}$ fields have identical
correlation functions \cite{dupuis1}.
\begin{figure}
\rotatebox{0}{\includegraphics*[width=\linewidth]{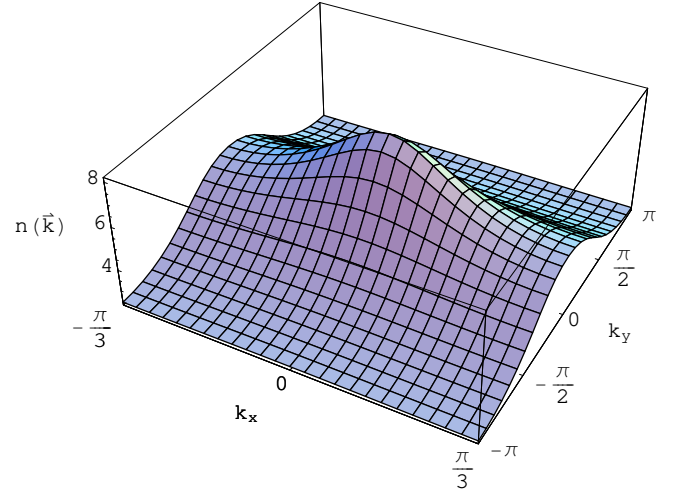}}
\caption{(Color online)Plot of $n({\bf k})$
 for $q=3$ and $p=1$ at $\mu/U=0.414$ and $t'/t'_c =.95$ showing peaks at
minima $(0,0)$.} \label{fig4}
\end{figure}
Next, we follow Refs.\ \onlinecite{dupuis1, sengupta1} to integrate
out the ${\tilde \psi}$ and $\phi$ fields and obtain an effective
action in terms of $\psi$. The details of this procedure has been
elaborated in Ref.\ \onlinecite{dupuis1}. The effective action so
obtained is given by \cite{dupuis1,sengupta1}
\begin{eqnarray}
{\mathcal S}_{\rm eff} &=& {\mathcal S}_0 + {\mathcal S}_1 \nonumber\\
{\mathcal S}_0 &=& \int_{\bf k} \, \psi_{q}^{\ast} (i\omega_n, {\bf
k}) [-G_0^{-1}(i\omega_n) I + \Lambda^q({\bf k}) ] \psi_{q} ( i
\omega_n,{\bf
k}), \label{s0eq} \nonumber\\
{\mathcal S}_1 &=& g/2 \int_0^{\beta} d \tau \int d^2 r |\psi_{q
}^{\ast}({\bf r},\tau) \psi_{q}({\bf r}, \tau)|^2, \label{s1eq}
\end{eqnarray}
where $\psi_q = (\psi_0 (k_x,k_y) .. \psi_{q-1}(k_x,k_y))^{T}$ with
$\psi_{\alpha}(k_x,k_y) = \psi(k_x+2 \pi \alpha/q,k_y)$ denoting the
$q$-component of the auxiliary field $\psi$ in momentum space, $\int
_{\bf k} \equiv (1/\beta) \sum_{\omega_n} \int d^2k/(2 \pi)^2$, $I$
denotes the unit matrix, and $g>0$ is the static limit of the exact
two-particle vertex function of the bosons in the local limit which
has been computed in Ref.\ \onlinecite{dupuis1}. Note that $S_0$
reproduces exact bosons propagator both in the local ($t'=0$) and
the non-interacting ($U=0$) limits and therefore provides a suitable
starting point for the strong coupling approximation.  In the next
subsection, we shall compute the momentum distribution function of
the bosons from $S_0$.

\subsection{ Momentum distribution of the bosons in the Mott phase}

The momentum distribution of the bosons in the Mott phase can be
computed from $S_0$ \cite{dupuis1,sengupta1}
\begin{eqnarray}
n({\bf k}) &=& -\lim_{T \to 0}(1/\beta) \sum_{\omega_n} {\rm Tr}
G(i \omega_n, {\bf k}), \nonumber\\
G(i\omega_n,{\bf k}) &=& [-G_0^{-1}(i\omega_n) I + \Lambda^q ({\bf
k};p) ]^{-1}.
\end{eqnarray}
\begin{figure}
\rotatebox{0}{\includegraphics*[width=\linewidth]{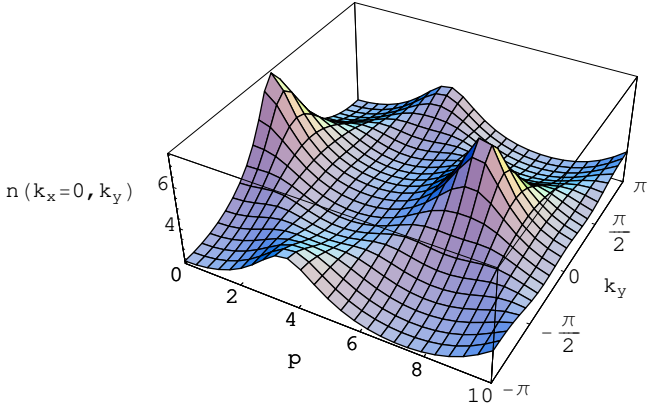}}
\caption{(Color online) Plot of $n(k_x=0,k_y)$ as a function of
$k_y$ and $p$ for $q=3$ at $\mu/U=0.414$ and
$t'/U=0.04$.}\label{fig5}
\end{figure}
To compute $n({\bf k})$, we note that $G_0^{-1}$ is independent of
momenta. Hence finding $G(i\omega_n,{\bf k})$ amounts to inverting
$\Lambda^q({\bf k};p)$. To this end we introduce an unitary
transformation where the transformation matrix $U_q({\bf k})$
diagonalizes $\Lambda^q({\bf k};p)$ to obtain a diagonal Green
function $G^d(i \omega_n, {\bf k})= U_q^{-1}({\bf k})G(i \omega_n,
{\bf k}) U_q({\bf k})$ whose diagonal elements are given by
\begin{eqnarray}
G^d_{\alpha \alpha}(i \omega_n, {\bf k})&=& [-G_0^{-1}(i\omega_n) +
\epsilon_{\alpha}^q ({\bf k};p)]^{-1} \nonumber\\
&=& \frac{i \omega_n + \mu + U}{(i \omega_n -E_{q}^{\alpha +} ({\bf
k};p))(i \omega_n -E_{q}^{\alpha -} ({\bf k};p))}, \nonumber\\
\label{greendiag}
\end{eqnarray}
where we have used the expression of $G_0$ from Eq.\
\ref{greenlocal} and $E^{q \pm}_{\alpha} ({\bf k};p)$ denote the
location of the poles of the interacting boson Green function and
are given by
\begin{eqnarray}
E_q^{\alpha \pm}({\bf k};p) &=& -\mu + U(n_0-1/2) +
\epsilon^q_{\alpha}({\bf k};p)/2 \pm \frac{1}{2} \nonumber\\
&& \times \sqrt{\epsilon^q_{\alpha}({\bf k};p)^2 + 4
\epsilon^q_{\alpha}({\bf k};p)U (n_0+1/2) + U^2}. \nonumber\\
\label{pole1}
\end{eqnarray}

Note that $E_q^{\alpha \pm}({\bf k};p)$ can be directly computed
from the knowledge of the non-interacting boson spectrum
$\epsilon^q_{\alpha}({\bf k};p)$ derived in Sec.\ \ref{sp}. In
particular, the minima $E_q^{\alpha \pm}({\bf k};p)$ occur in the
same position in the magnetic Brillouin zone as
$\epsilon^q_{\alpha}({\bf k};p)$. Also, as noted in Ref.\
\onlinecite{dupuis1}, the Mott gap $E_q^{\alpha +}({\bf k};p)-
E_q^{\alpha -}({\bf k};p)$ vanishes at the position of the minima of
$\epsilon^q_{\alpha}({\bf k};p)$ in the magnetic Brillouin zone
provided we are at the tip of the Mott lobe where the SI transition
takes place at constant density.

\begin{figure}
\rotatebox{0}{\includegraphics*[width=\linewidth]{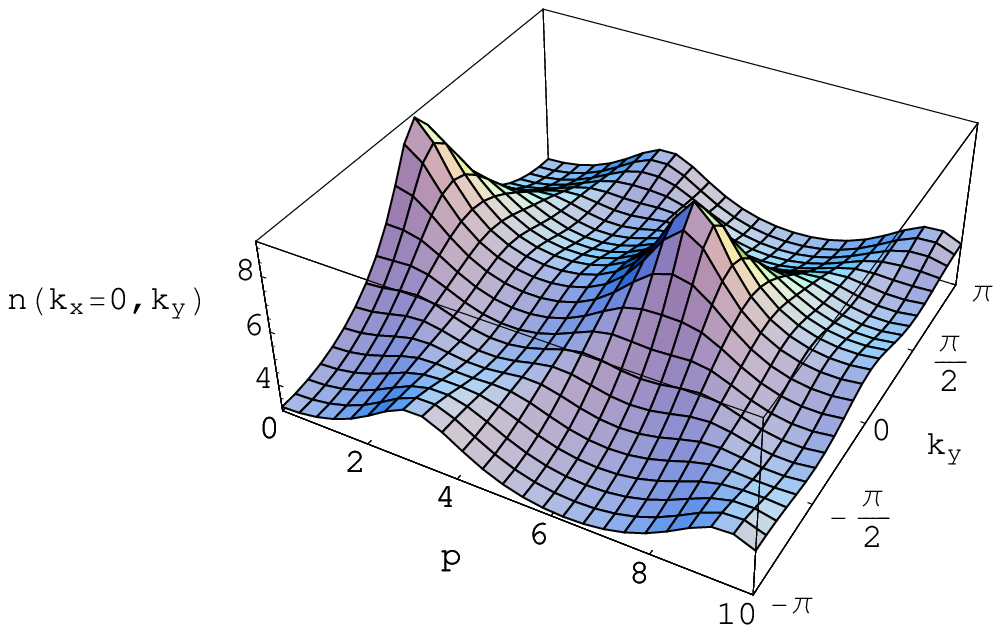}}
\caption{(Color online) Same as Fig.\ \ref{fig5} for $q=4$.}
\label{fig6}
\end{figure}

\begin{figure}
\rotatebox{0}{\includegraphics*[width=\linewidth]{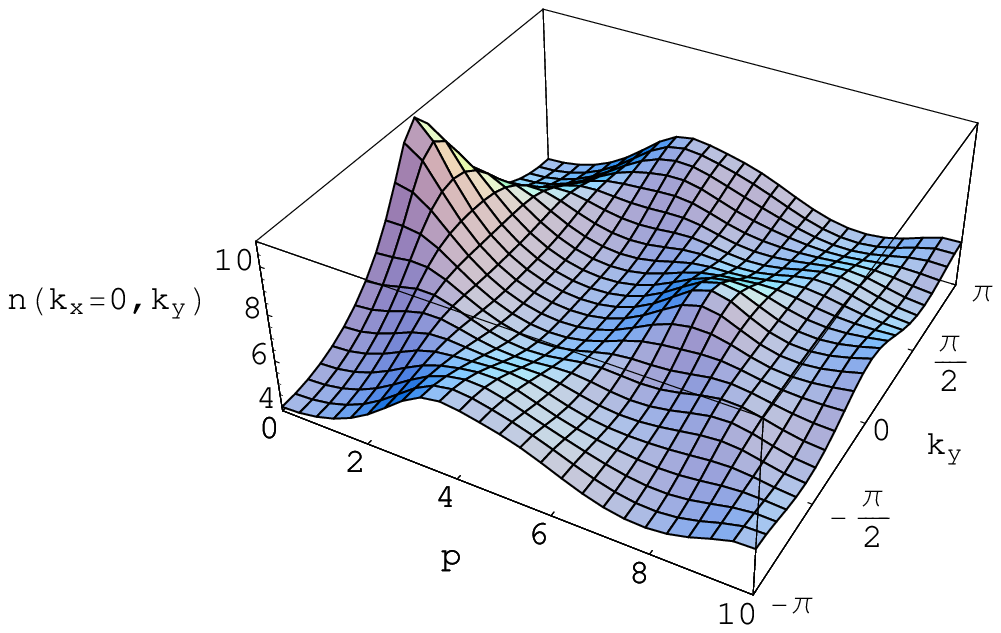}}
\caption{(Color online)Same as Fig.\ \ref{fig5} for $q=5$.}
\label{fig7}
\end{figure}
The momentum distribution can now be computed as $n({\bf k})=
-\lim_{T \to 0}(1/\beta) \sum_{\omega_n} {\rm Tr} G^d(i \omega_n,
{\bf k})$ and is given by \cite{sengupta1}
\begin{eqnarray}
n({\bf k}) &=& \sum_{\alpha=0}^{q-1} \frac{E_q^{\alpha -}({\bf
k};p)+\mu +U}{E_q^{\alpha +}({\bf k};p)- E_q^{\alpha -}({\bf k};p)}.
\label{momdist1}
\end{eqnarray}
Eq.\ \ref{momdist1} shows that the peaks of $n({\bf k})$ occur when
the Mott gap $E_q^{\alpha +}({\bf k})-E_q^{\alpha -}({\bf k})$
becomes small near the minima of $\epsilon^q_{\alpha}({\bf k};p)$ as
the SI transition is approached through the tip of the Mott lobe.
The minima structure of the non-interacting bosons is therefore
expected to be reflected in the peaks of the momentum distribution
of the bosons in the Mott phase. In Fig.\ \ref{fig4}, we show a
representative plot of $n({\bf k})$ as a function of ${\bf k}$ for
$q=3$ and $p=1$. We find that the central peak of the momentum
distribution lies at $(0,0)$ in accordance with the position of the
minima of $\epsilon^{(3)}_0({\bf k},1)$. Next, keeping in mind that
the position of the minima of $\epsilon^q_{\alpha}({\bf k};p)$
always occur at $k_x=0$, we plot the momentum distribution
$n(k_x=0,k_y)$ as a function of $k_y$ (for fixed $t'(p)/t'_c(p)
=0.95$ and $q=3,5$) for several representative values of $p$ in Fig.
\ref{fig3}. Fig.\ \ref{fig3} clearly shows that as $p$ increases,
the peak structure of the momentum distribution changes from a
single peak at $k_y=0$ to two split peaks at $k_y=\pm k_y^{\rm
min}(p)$ and finally to a single peak at $k_y=\pi$. Finally in
Figs.\ \ref{fig5}, \ref{fig6} and \ref{fig7}, we plot $n(k_x=0,k_y)$
for $q=3$, $4$, and $5$, as a function of $k_y$ and $p$ for a fixed
$t'=0.04U$. Note that for these plots, the proximity of the system
to the tip of the Mott lobe changes with $p$ since $t'_c$ is a
function of $p$. These plots again reveal the change in the peak
structure of $n(k_x=0,k_y)$ as a function of $p$.

\subsection{Re-entrant SI transitions}

The critical hopping $t'_c$ for the MI-SF transition as a function
of $\mu$ can be determined from the condition \cite{sengupta1}
\begin{eqnarray}
r_q(p) &=& -G_0^{-1}(i\omega_n=0)+\epsilon^q_{{\rm min}}(p)= 0.
\label{transcond1}
\end{eqnarray}
\begin{figure}
\rotatebox{0}{\includegraphics*[width=\linewidth]{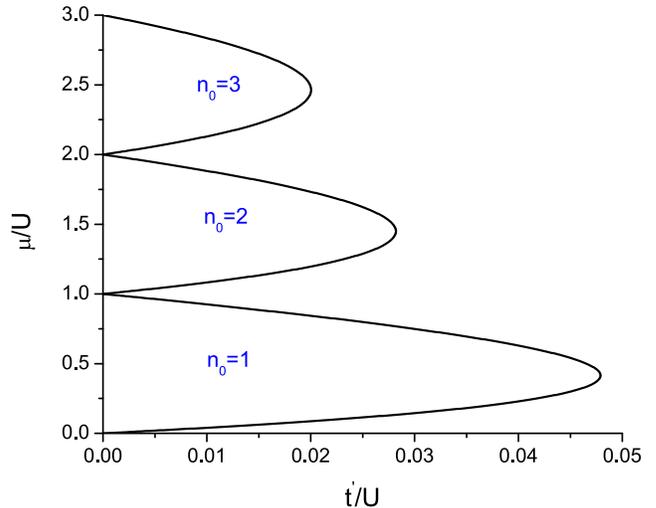}}
\caption{(Color online)The MI-SF phase boundary for $q=3$ and
$p=1$.} \label{fig8}
\end{figure}
The SI phase boundary so obtained is shown in Fig.\ \ref{fig1} for
$q=3$ and $p=1$ in Fig.\ \ref{fig8} and displays the usual Mott
lobes. The difference of the present case here with the SI
transitions studied earlier \cite{fisher1,dupuis1,
sengupta1,freericks1} arises due to the non-monotonic $p$ dependence
of $\epsilon^q_{{\rm min}}(p)$. This point is demonstrated in Fig.\
\ref{fig9} for $q=3$, $4$, $5$, and $6$ by plotting $t'_c(p)$ as a
function of $p$ for $n_0=1$ and $\mu=\mu_{\rm tip}$. We find that
$t'_c(p)$ is a non-monotonic function of $p$ and $t'_c(p) > t'_c(0)$
for all $p$, Consequently, varying $p$ at a fixed value of
$t'>t'_c(0)$ leads to a series of field-induced re-entrant SI
transitions for any $q$. This is schematically marked by the
red-dotted line in Fig.\ \ref{fig9}. We note that such re-entrant
transitions as a function of the magnetic field strength are not
present for SI transitions in a constant magnetic field
\cite{sengupta1}.

\section{The superfluid phase}
\label{sf}

At $t'=t'_c(p)$, it becomes energetically favorable to create
particles/holes at the minima of the energy dispersion of the bosons
leading to the destabilization of the Mott phase. The
Landau-Ginzburg theory of the resultant superfluid phase can be
expressed by long-wavelength boson fields around these minima. In
the present case, there are either one or two degenerate minima of
the boson energy spectrum in the magnetic Brillouin zone leading to
a Landau-Ginzburg theory of one or two low-energy boson fields
\cite{fisher1,dupuis1,sengupta1,freericks1}.
\begin{figure}
\rotatebox{0}{\includegraphics*[width=\linewidth]{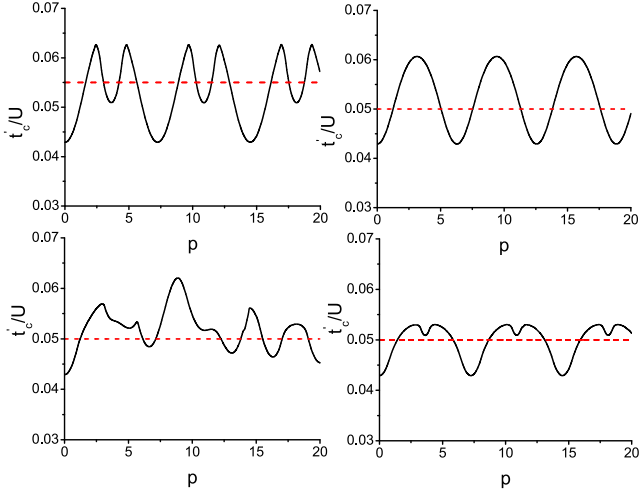}}
\caption{(Color online) Plot of the critical hopping strength
$t'_c(p)$ as a function of $p$ at the tip of the Mott lobe ($\mu=
\mu_{\rm tip}=0.414U$) for $q=3$ (left top panel), $4$ (right top
panel), $5$ (left bottom panel), and $6$ (right bottom panel). The
red-dashed line is a guide to the eye showing reentrant SI
transitions as $p$ is varied at fixed $t' > t'_c(p=0)$.}\label{fig9}
\end{figure}
We shall first consider the case with a single minima either at
$(0,0)$ or $(0,\pi)$ which occurs for specific ranges of $p$ for all
$q$ as discussed in Sec.\ \ref{sp}. In either case, the boson field
can be written as
\begin{eqnarray}
\psi({\bf r},t) &=& \chi_0({\bf r};p)\varphi({\bf r},t), \nonumber\\
\chi_0({\bf r};p) &=& \left[\sum_{\alpha=0}^{q-1} \psi_{\alpha}(p)
e^{2 \pi i \alpha x/q} \right] e^{ik_y^{\rm min} y} \varphi({\bf
r},t), \label{bosonfield}
\end{eqnarray}
where $\psi_{\alpha}(p)$ denotes the components of eigenvectors of
$\Lambda_q({\bf k};p)$ at $k_x=0$, $ k_y=k_y^{\rm min}$ which can be
either $0$ or $\pi$ for a fixed $p$, and $\chi_0({\bf r};p)$ denotes
the corresponding wavefunction in real space. Thus the superfluid
density can be written as
\begin{eqnarray}
\rho_s({\bf r}) &=& \left|\langle \psi \rangle \right|^2 =
\left|\sum_{\alpha=0}^{q-1} \psi_{\alpha}(p) e^{2 \pi i \alpha x/q}
\right|^2 |\varphi_0|^2, \label{supfleq1}
\end{eqnarray}
where $\varphi_0 = \langle \varphi({\bf r},t) \rangle \ne 0$ for $t'
> t'_c(p)$. Note that $\rho_s$ is independent of $y$ irrespective of
the value of $k_y^{\rm min}$, but displays spatial variation along
$x$. Further, as discussed in Sec.\ \ref{sp}, for even $q$, only
$q/2$ of the components $\psi_{\alpha}$ (corresponding to either
even or odd integers $\alpha$) will be non-zero. Consequently, we
expect the period of $\rho_s(x)$ to be halved. A plot of the
renormalized superfluid density $\rho_s(x)/\rho_s(0)$, plotted in
Fig.\ \ref{fig10} for $p=0.5$ and $q=3$, $4$, $5$, and $6$, confirms
this expectation. The presence of the periodic vector potential
leads to a $q$- periodic pattern with $q-2$ small and one large peak
in the superfluid density along $x$ for all odd $q$ as shown in the
left panels of Fig.\ \ref{fig10}. In contrast, the superfluid
density for even $q$ displays a $q/2$ periodic pattern. Note that
this period halving leads to identical superfluid density patterns
for vector potentials with periods $q$ and $2q$ for all odd $q$.
This feature is clearly demonstrated in the top left ($q=3$) and the
bottom right ($q=6$) panels of Fig.\ \ref{fig10}.

\begin{figure}
\rotatebox{0}{\includegraphics*[width=\linewidth]{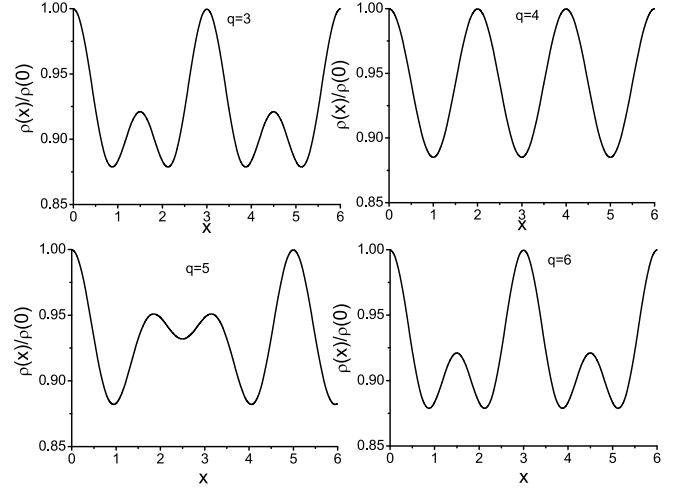}}
\caption{(Color online) Plot of the superfluid density
$\rho_s(x)/\rho_s(0)$ as a function of $x$ for $q=3,5 $ (left
panels) and $q=4,6$ (right panels). Note that the superfluid density
displays a $q$ periodic pattern for odd $q$s and a $q/2$ periodic
pattern for even $q$s. $p$ is set to $0.5$ for all
plots.}\label{fig10}
\end{figure}

Next, we derive the effective low-energy Landau-Ginzburg theory. To
this end, we substitute Eq.\ \ref{bosonfield} in Eq.\ \ref{s1eq} and
obtain the effective low-energy Landau-Ginzburg action in terms of
the $\varphi$ fields. The details of this procedure is charted out
in Ref.\ \onlinecite{sengupta1}. The resultant action is given by
\begin{eqnarray}
S^{\rm LG}_1 &=& \int d^2 r dt \Big[\varphi^{\ast}({\bf r},t) \big[
K_0
\partial_{t}^2 + i K_1
\partial_{t} +r_q (p) \nonumber\\
&&  - v_q(p) ^2 (\partial_x ^2 + \partial_y^2) \big] \varphi({\bf
r},t) + \frac{g'}{2} |\varphi({\bf r},t)|^4 \Big], \label{lgsingle}
\end{eqnarray}
where $K_0 = 1/2 \partial^2 G_0^{-1}/\partial \omega^2|_{\omega=0} =
n_0(n_0+1)U^2/(\mu+U)^3$, $K_1=
\partial G_0^{-1}/\partial \omega|_{\omega=0} = 1 -
n_0(n_0+1)U^2/(\mu+U)^2$, and $v_q(p)^2 = \nabla^2_{{\bf k}}
\epsilon_{\rm min}({\bf k};p)/2$, $r_q (p)$ is given by Eq.\
\ref{transcond1}, and $g' = g \sum_{x,y=0}^{q-1} |\chi_0({\bf
r};p)|^4/q^2$. At the tip of the Mott lobe, where $\mu= \mu_{\rm
tip} =U(\sqrt{n_0(n_0+1)}-1)$, $K_1=0$. Thus we have a critical
theory with dynamical critical exponent $z=1$. Away from the tip,
$K_1 \ne 0$ rendering $z=2$. Thus the critical theory turns out to
have similar exponent as in the case without magnetic field
\cite{sachdev1}.

\begin{figure}
\rotatebox{0}{\includegraphics*[width=\linewidth]{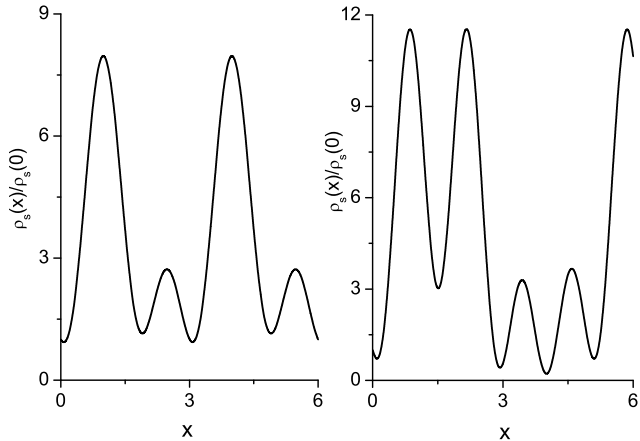}}
\caption{(Color online) Same plot as Fig.\ \ref{fig10}
 for $q=3$ ($k_y^{\rm min}= 2.2)$) (left panel)
 and $q=5$ ($k_y^{\rm min}=1.89$)
(right panel) showing $q$ periodic patterns. $p=2.5$ for both cases.
}\label{fig11}
\end{figure}

Finally, we briefly comment on the case where there are two
degenerate minima either at $(0,\pm k_y^{\rm min})$ or at $(0,0)$
and $(0,\pi)$. In this case, $\psi({\bf r},t)= \chi_0^+({\bf
r};p)\varphi_+({\bf r},t)+\chi_0^-({\bf r};p)\varphi_-({\bf r},t)$
where $\chi_0^{\pm}({\bf r})$ denotes the eigenfunctions of
$\Lambda({\bf k};p)$ in real space at $(0,\pm k_y^{\rm min})$ and
$\varphi_{\pm}({\bf r},t)$ denotes low-energy fluctuating fields
about the  minima. Substituting this expression of $\psi$ in Eq.\
\ref{s1eq}, and following the coarse-graining procedure detailed in
Ref.\ \onlinecite{sengupta1}, we find that for all $q$ and $p$, the
superfluid phase corresponds to the condensation of only one of the
low-energy fields: $\langle \varphi_+ \rangle=0, \, \langle
\varphi_-\rangle\neq0$ or $\langle \varphi_- \rangle=0, \, \langle
\varphi_+\rangle\neq0$. Thus the effective Landau-Ginzburg action in
these cases is qualitatively similar to Eq.\ \ref{lgsingle}. The
superfluid density, plotted in Fig.\ \ref{fig11} for $q=3,5$ and
$p=2.5$, shows similar $q$ periodic pattern as observed in Fig.\
\ref{fig10} for odd $q$.

\section{Discussion}
\label{conc}

There are several possible experimental verifications of our theory.
First, we suggest measurement of $n({\bf k})$ for the bosons in the
Mott phase near the transition as done earlier in Ref.\
\onlinecite{spielman1} for 2D optical lattices without the synthetic
magnetic field. Our prediction is that the peak structure of the
momentum distribution along $k_x=0$ at a fixed $t'/U$ near $t'_c$
would be similar to those shown in Figs.\ \ref{fig5}..\ref{fig7}. In
particular the shift in the peak position of $n(0,k_y)$ with $p$ and
change from a single to double peak structure as a function of $p$
should be observable in such experiments. Second, the re-entrant SI
transition can also be verified by measuring $n({\bf k})$ as a
function of $p$ by fixing $t'>t'_c(p=0)$ as shown in Fig.\
\ref{fig9}. Finally, the spatial variation of the superfluid density
can also be observed by measuring $n({\bf k})$ in the superfluid
phase.

In conclusion, we have analyzed the MI-SF transition of ultracold
bosons in a 2D optical lattice in the presence of a synthetic
periodic magnetic field.  We have shown that the precursor peaks of
the momentum distribution in the Mott phases can be tuned by the
strength $p$ of the synthetic field. We have also demonstrated that
the bosons, in the presence of such a periodic synthetic magnetic
field, show a series of field-induced re-entrant SI transitions, and
that the superfluid density in the SF phase near criticality shows
$q$ ($q/2$) periodic spatial pattern for odd (even) $q$. We have
suggested several experiments which can test our theory.

K.S. thanks R. Shankar for discussions and DST, India for financial
support under Project No. SR/S2/CMP-001/2009.

\end{document}